\documentclass[12pt]{article}
\usepackage{graphicx}
\pdfoutput=1

\newcommand\pubnumber{DPF2013-256}
\newcommand\pubdate{\today}

\def\napoli{Department of Physics\\
Colorado State University, Fort Collins, CO, USA}

\def\Title#1{\begin{center} {\Large #1 } \end{center}}
\def\Author#1{\begin{center}{ \sc #1} \end{center}}
\def\Address#1{\begin{center}{ \it #1} \end{center}}

\newcommand\pubblock{\rightline{\begin{tabular}{l} \pubnumber\\
         \pubdate  \end{tabular}}}
\newenvironment{Abstract}{\begin{quotation}  }{\end{quotation}}
\newenvironment{Presented}{\begin{quotation} \begin{center} 
             PRESENTED AT\end{center}\bigskip 
      \begin{center}\begin{large}}{\end{large}\end{center} \end{quotation}}

\def\beq{\begin{equation}}
\def\eeq#1{\label{#1}\end{equation}}
\def\eeqn{\end{equation}}

\def\beqa{\begin{eqnarray}}
\def\eeqa#1{\label{#1}\end{eqnarray}}
\def\eeqan{\end{eqnarray}}
\def\CR{\nonumber \\ }

\let\bar=\overbar

\def\Dslash{\not{\hbox{\kern-4pt $D$}}}
\def\dslash{\not{\hbox{\kern-2pt $\del$}}}

\def\msb{{\bar{\ssstyle M \kern -1pt S}}}

\begin{document}
\begin{titlepage}
\pubblock

\vfill
\Title{Future Neutrino Oscillation Sensitivities for LBNE}
\vfill
\Author{Matthew Bass\footnote{Presented by Matthew Bass at the DPF 2013 Meeting of the American Physical Society
Division of Particles and Fields, Santa Cruz, California, August 13-17, 2013}, Daniel Cherdack and Robert J. Wilson\\For the LBNE Collaboration}
\Address{\napoli}
\vfill
\begin{Abstract}
The primary goal of the Long-Baseline Neutrino Experiment (LBNE) is to measure the neutrino mixing matrix parameters. The design, optimized to search for CP violation and to determine the neutrino mass hierarchy, includes a large $\mathcal{O}(10$ kt) Liquid Argon Time Projection Chamber (LAr TPC) at 1300 km downstream of a wide-band neutrino beam. A brief introduction to the neutrino mixing parameters will be followed by a discussion of sensitivity study analysis methods and a summary of the results for LBNE. The studies include comparisons with the Tokai-to-Kamioka (T2K) and NuMI Off-axis electron-neutrino Appearance (NO$\nu$A) experiments as well as combined sensitivities. Finally, the impact of including a realistic set of systematic uncertainties will be presented.
\end{Abstract}
\vfill
\begin{Presented}
DPF 2013\\
The Meeting of the American Physical Society\\
Division of Particles and Fields\\
Santa Cruz, California, August 13--17, 2013\\
\end{Presented}
\vfill
\end{titlepage}
\def\thefootnote{\fnsymbol{footnote}}
\setcounter{footnote}{0}

\section{Introduction}
Neutrino flavor oscillation, or mixing, has been well established experimentally. However the parameters that describe this mixing for the three-neutrino scenario, the elements of the PMNS matrix\cite{Pontecorvo:1957cp}\cite{Pontecorvo:1967fh}\cite{Maki:1962mu}, have been measured to various levels of precision. These parameters consist of three angles $\theta_{12}$, $\theta_{13}$, and $\theta_{23}$, and Charge-Parity (CP) violating phase, $\delta_{CP}$, that govern the amplitude of the mixing. The frequency of the oscillations with respect to baseline and neutrino energy is determined by the mass squared differences: $\Delta m^2_{21}$, $\Delta m^2_{31}$, and $\Delta m^2_{32}$, of which two are distinct. The angles $\theta_{12}$ and $\theta_{23}$ as well as the mass squared differences have been well constrained by several neutrino experiments. The angle $\theta_{13}$ has recently been constrained by accelerator based experiments that have measured $\nu_\mu\to\nu_e$ appearance rates and reactor experiments that have measured electron-neutrino disappearance rates.

There are, however, still open questions to be answered. The value of $\delta_{CP}$ is currently not well constrained by any experiment. A non-zero value of $\sin \delta_{CP}$, signifying a CP violating process, would mean that neutrinos and anti-neutrinos mix differently. 
Although current experimental data is consistent with maximal mixing for $\theta_{23}$, if more precise measurements find $\theta_{23}$ to be less than maximal then $\nu_\mu\to\nu_e$ oscillations will be sensitive to the $\theta_{23}$ octant.
Finally, while the values of $\Delta m^2_{21}$ and $|\Delta m^2_{31}|$ are known\cite{Fogli:2012ua} within 3\%, the sign of $|\Delta m^2_{31}|$, or the neutrino mass hierarchy, is unknown. A positive(negative) value is denoted the normal(inverted) hierarchy.

\begin{figure}[htb]
\centering
\includegraphics[width=0.3553\textwidth,trim=0in 0in 0.25in 0in,clip]{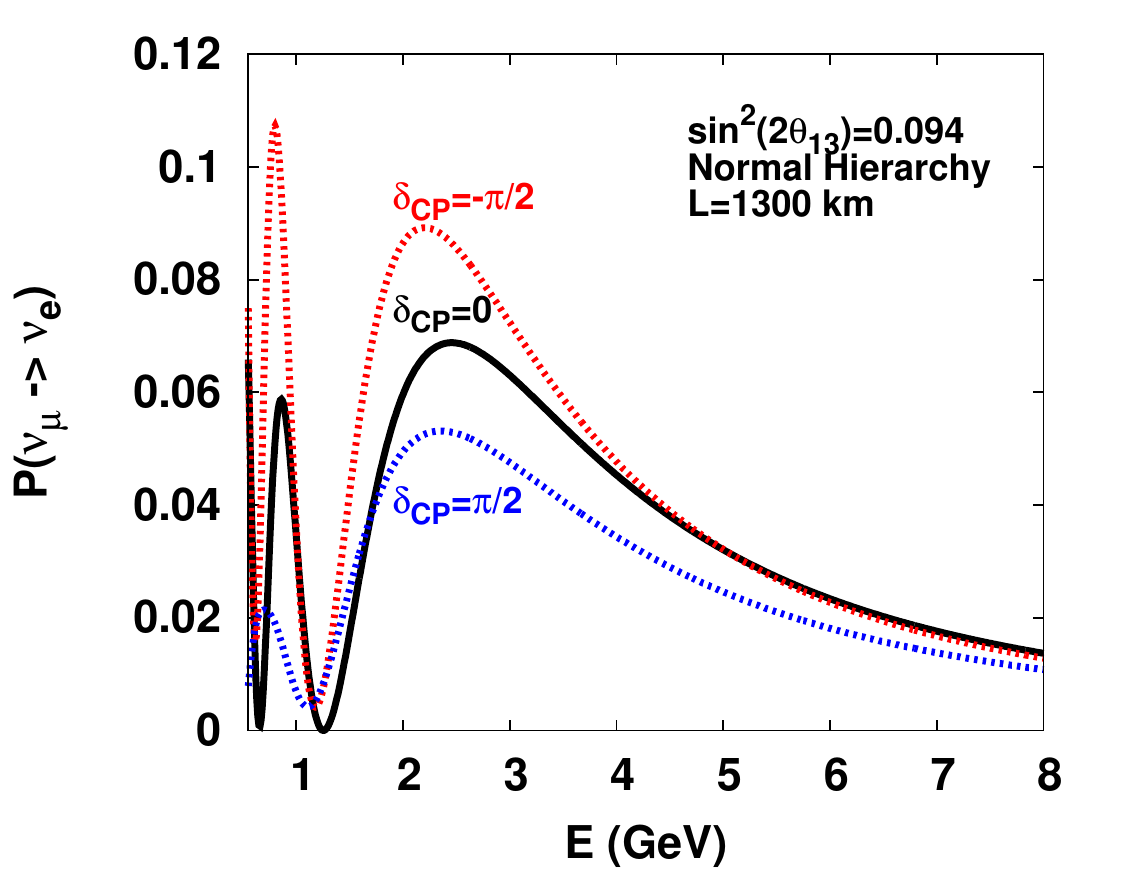}
\includegraphics[width=0.28\textwidth,trim=0.9in 0in 0.25in 0in,clip]{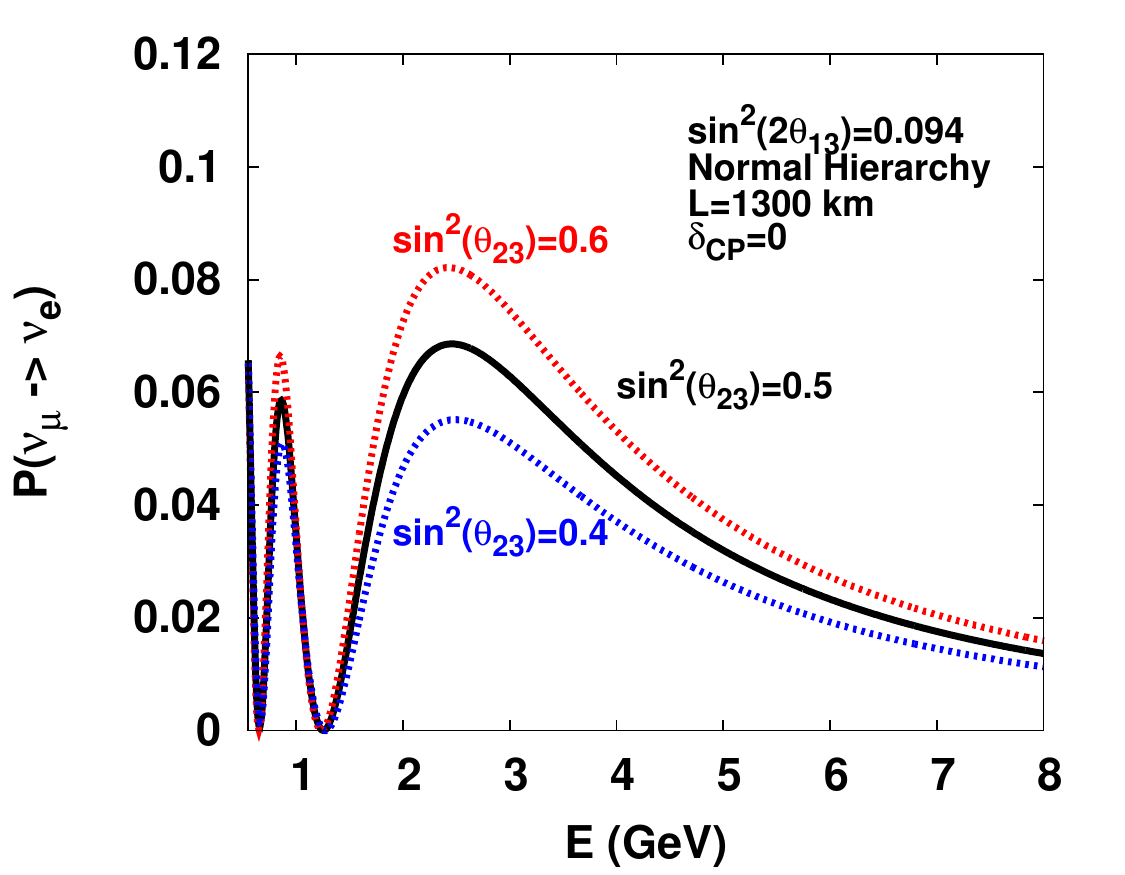}
\includegraphics[width=0.28\textwidth,trim=0.9in 0in 0.25in 0in,clip]{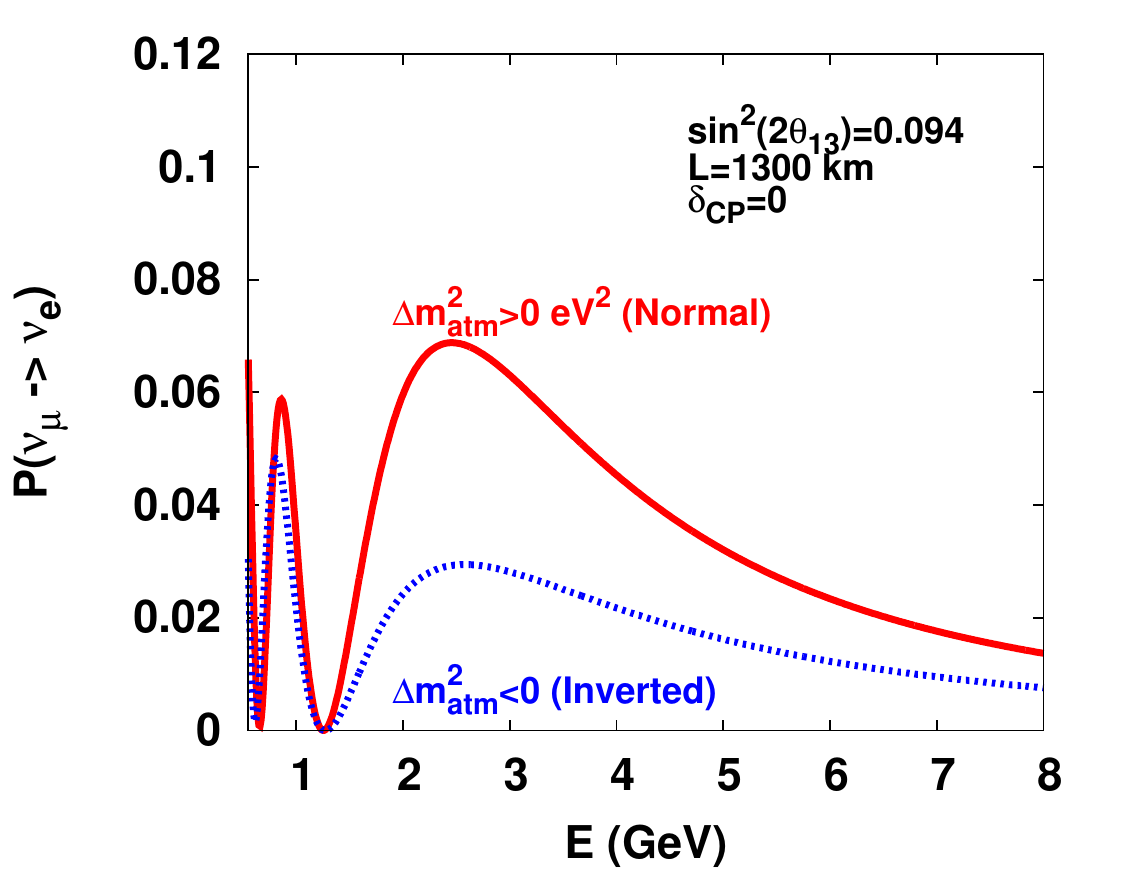}
\caption{Probabilities for $\nu_\mu\to\nu_e$ as a function of neutrino energy for variations in $\delta_{CP}$ (left), $\sin^2\theta_{23}$ (middle), and mass hierarchy (right) at 1300 km.}
\label{fig:probs}
\end{figure}

An approximate form\cite  {Freund:1999gy} for the probability of $\nu_\mu\to\nu_e$ in matter is

\beqa
  P_{{\nu_{\mu}}\rightarrow{\nu_{e}}} &\approx& \sin^2 2\theta_{13} \sin^2\theta_{23}
  \frac{\sin^2((1-x)\Delta)}{\left(1-x\right)^2} \nonumber \CR 
  &-& \alpha \sin2\theta_{13} \sin \delta \sin 2\theta_{12} \sin 2\theta_{23}
  \sin \Delta \frac{\sin(x \Delta)}{x} \frac{\sin((1-x)\Delta)}{\left(1-x\right)} \nonumber \CR
  &+& \alpha \sin2\theta_{13} \cos \delta \sin 2\theta_{12} \sin 2\theta_{23}
  \cos \Delta \frac{\sin(x \Delta)}{x} \frac{\sin((1-x)\Delta)}{\left(1-x\right)} \nonumber \CR
  &+& \alpha^2 \cos^2 \theta_{23} \sin^2 2\theta_{12} \frac{\sin^2(x\Delta)}{x^2},
  \label{eqn:osc}
\eeqan

where $x=\frac{2\sqrt{2}G_F N_e E}{\Delta m^2_{31}}$, $\Delta=\frac{\Delta m^2_{31} L}{4 E}$, and $\alpha=\frac{\Delta m^2_{21}}{\Delta m^2_{31}}\approx0.03$.
All six of the parameters governing neutrino oscillations appear in this equation. 
The current constraints on the oscillation parameters still allow for functionally degenerate solutions to Eq. \ref{eqn:osc}.
Figure \ref{fig:probs} shows schematically how these degeneracies arise at a baseline (L) of 1300 km by plotting the probabilities for $\nu_e$ appearance as a function of neutrino energy. For example, a variation of $\delta_{CP}$ from 0 to $-\pi/2$ resembles a variation of $\sin^2\theta_{23}$ from 0.5 to 0.6. So there is a functional degeneracy between the two parameters. An experiment with sufficient energy resolution, statistics, and understanding of systematic uncertainties will be required to disentangle their effects.

\section{LBNE}
The LBNE\cite{Adams:2013qkq} is currently designed with a 700 kW wide-band muon-neutrino beam from the Fermi National Accelerator Laboratory (FNAL) to the Sanford Underground Research Facility (SURF). At a baseline of 1300 km from the beam source, a massive liquid argon TPC will be built underground. Also planned is a near detector, at FNAL, capable of reducing systematics to levels comparable with the assumptions used in these studies. 

LBNE is planning for a staged construction. In one scenario, LBNE10 will have a 10 kt fiducial mass LAr TPC with a 700 kW, 120 GeV beam. The proposed LBNE has a 34 kt fiducial mass LAr TPC, underground with a 700-2300 kW beam (see Project X\cite{Kronfeld:2013uoa}) in addition to a near neutrino detector at FNAL.

With its optimized 1300 km baseline, LBNE will further constrain $\theta_{23}$, determine the $\theta_{23}$ octant (for $\theta_{23} \neq 45^\circ$), determine the mass hierarchy, and detect CP violation for previously unreachable regions of the neutrino oscillation parameter phase space. This work focuses on the physics goals obtainable through $\nu_e$ appearance and $\nu_\mu$ disappearance analyses. There are many other physics goals for LBNE which are outlined in \cite{Adams:2013qkq}.
 
\section{Method for Estimating Sensitivity}
Sensitivities are computed using the GLoBES\cite{Huber:2004ka}\cite{Huber:2007ji} library. Event spectra for LBNE and LBNE10 are simulated using parameterizations\cite{Adams:2013qkq} of the flux, cross sections, energy resolutions, and analysis sample selection efficiencies. Event spectra for $\nu$ and $\bar{\nu}$ modes in both $\nu_e$ appearance and $\nu_\mu$ disappearance are considered. A $\Delta \chi^2$ is computed comparing a true event spectrum with test hypothesis spectra. The $\Delta \chi^2$ is minimized with respect to oscillation parameter uncertainties, adapted from the Fogli et al. 2012 global fit\cite{Fogli:2012ua}, and normalization uncertainties of 1\% on signal and 5\% on background events. For these studies, LBNE and LBNE10 are assumed to run with equal periods of $\nu$ and $\bar{\nu}$ running.

\section{LBNE Sensitivity}
\begin{figure}[!h]
  \centering
  \includegraphics[width=0.49\textwidth,trim=0in 0.221in 0in 0.4in,clip]{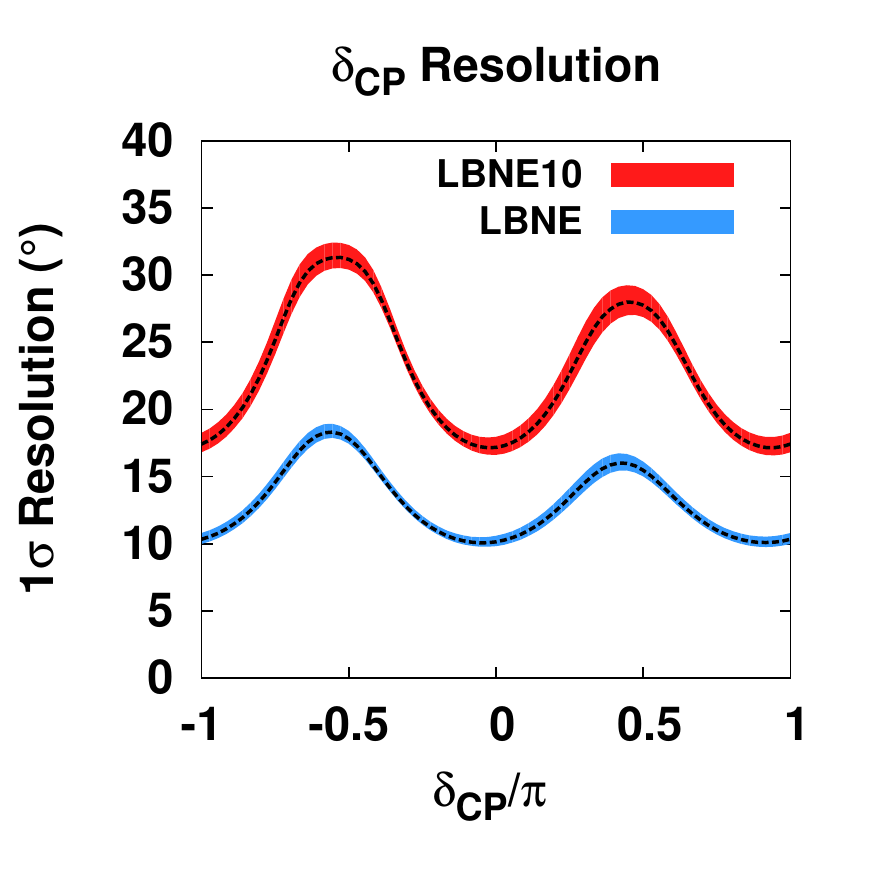}
  \includegraphics[width=0.455\textwidth,trim=0in 0in 0in 0.4in,clip]{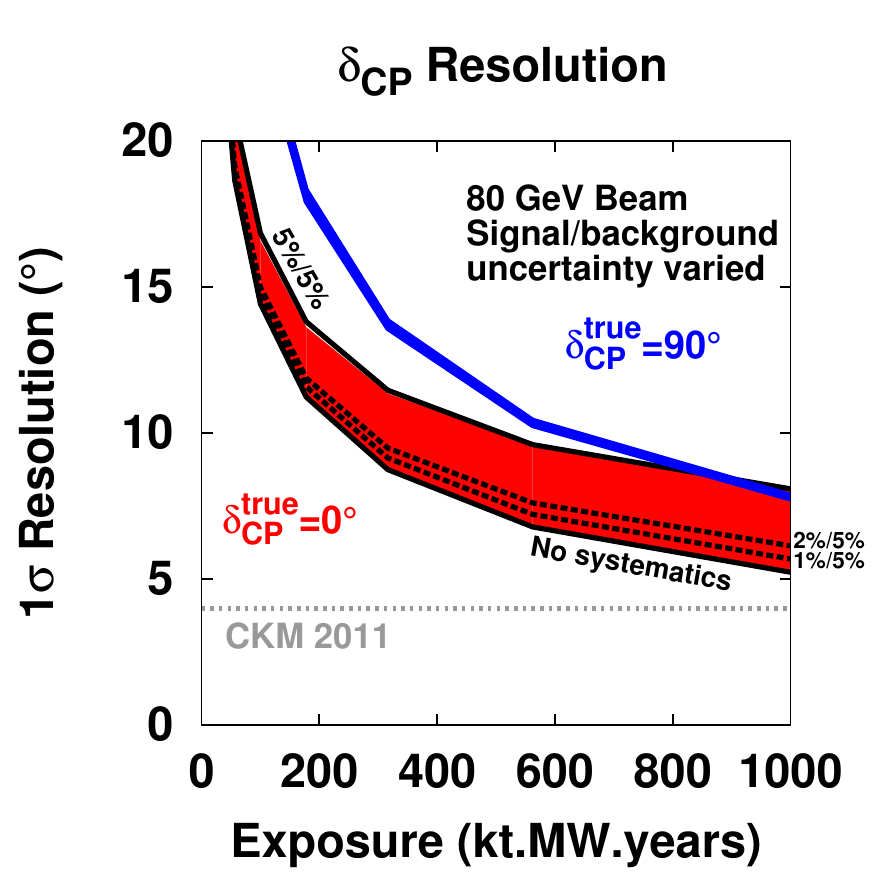}
  \caption{(left) Expected $1\sigma$ resolution on $\delta_{CP}$ for LBNE10 and LBNE. The bands represent current uncertainties on $\theta_{23}$ and $\Delta m^2_{31}$ and projected uncertainties on $\theta_{13}$ from the Daya Bay experiment. The mass hierarchy is assumed to be known and normal. (right) Expected $1\sigma$ resolution on $\delta_{CP}$ as a function of exposure for $\delta_{CP}=0^\circ$ (red) and $\delta_{CP}=90^\circ$ (blue) for multiple assumptions on signal and background normalization uncertainties. The level of precision of measurements for the CKM matrix describing quark mixing is given for comparison. }
  \label{fig:dcpres}
\end{figure}

Figure \ref{fig:dcpres} gives the expected resolution on $\delta_{CP}$, in degrees, as a function of the true value of $\delta_{CP}$ where the mass hierarchy is assumed to be known to be the normal hierarchy. LBNE10 will measure $\delta_{CP}$ to $\sim17^\circ$ in the best case and $\sim31^\circ$ in the worst case. The full LBNE, at 34 kt, will be able to measure the $\delta_{CP}$ resolution to between $10^\circ$ and $17^\circ$. The right panel of Fig. \ref{fig:dcpres} shows how these resolutions improve for two values of $\delta_{CP}^{true}$ as a function of exposure in kt$\cdot$MW$\cdot$years and for various uncertainties on signal and background normalizations. Uncertainties in the shape of the reconstructed energy spectra are not considered for these studies. These results depend upon the assumptions made concerning the systematic uncertainties on signal and background. The effect of varying these uncertainties can be seen in Fig. \ref{fig:dcpres}. More detailed treatments of systematic uncertainties are being considered within the context of the LBNE Fast Monte Carlo simulation\cite{Adams:2013qkq}.

\begin{figure}[!h]
  \centering
  \includegraphics[width=0.4\textwidth,trim=0.14in 0in 0.4in 0.4in,clip]{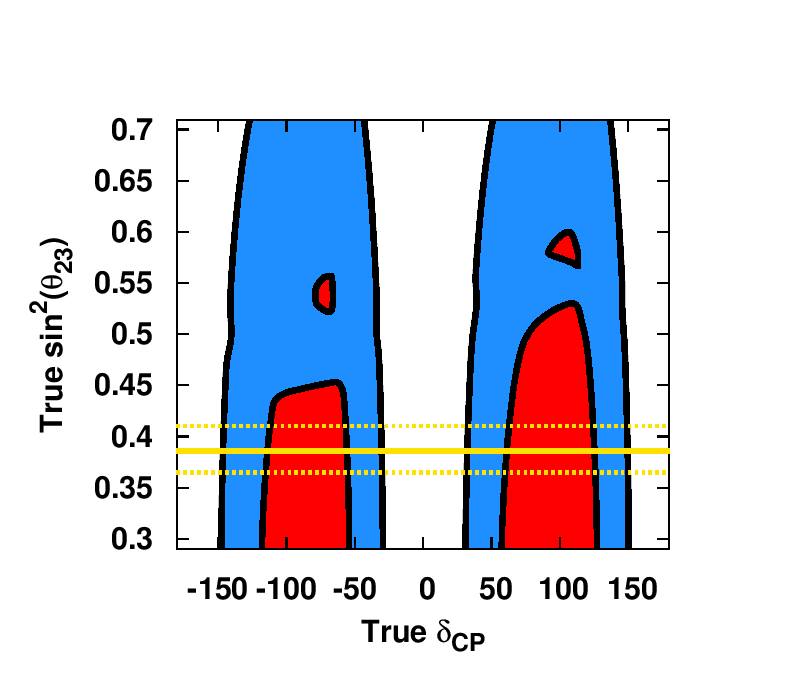}
  \includegraphics[width=0.4\textwidth,trim=0in 0in 0in 0.4in,clip]{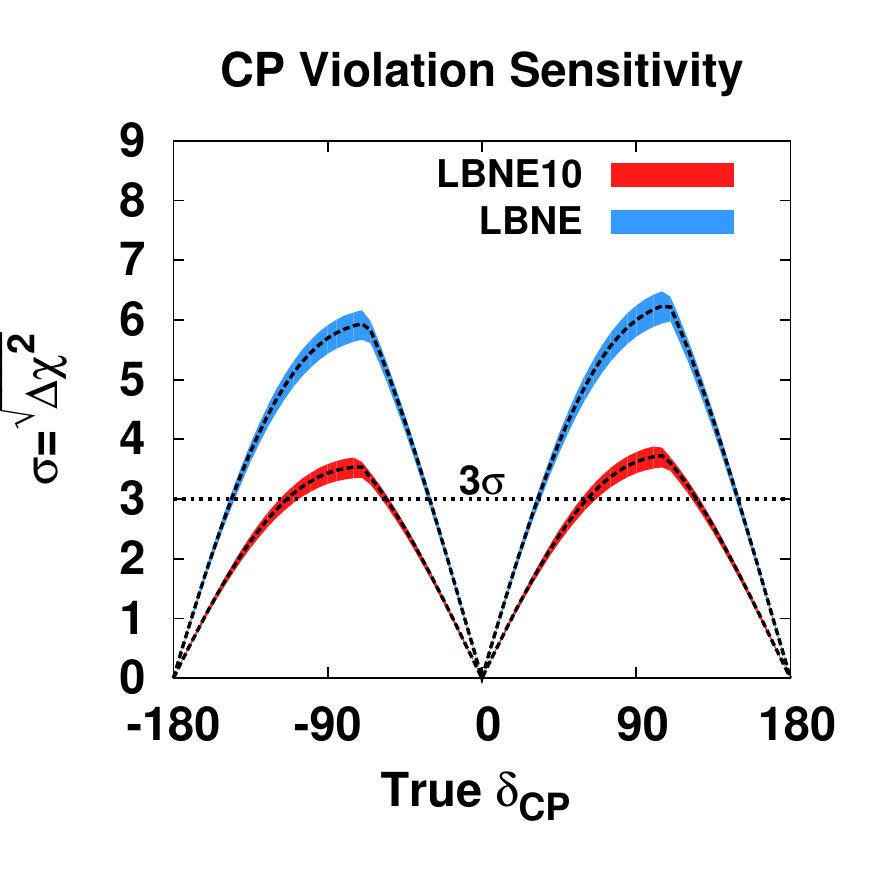}
  \caption{CP violation sensitivity assuming true normal hierarchy. (left) Regions in $\sin^2\theta_{23}$ vs. $\delta_{CP}$ where LBNE (cyan) and LBNE10 (red) have $3\sigma$ sensitivity to CP violation. The central value($1\sigma$ range) on $\sin^2\theta_{23}$ from the Fogli et al. 2012 global fit are indicated with the solid(dashed) yellow lines. (right) Expected sensitivity, in $\sigma$, with which LBNE and LBNE10 will be able to determine CP violation at $\sin^2\theta_{23}=0.39$. The bands represent current uncertainties on $\theta_{23}$ and $\Delta m^2_{31}$ and projected uncertainties on $\theta_{13}$.}
  \label{fig:cpv} 
\end{figure}

Figure \ref{fig:cpv} illustrates the sensitivity of LBNE10 and LBNE to CP violation. The left panel shows the values of $\sin^2\theta_{23}$ and $\delta_{CP}$ for which CP violation can be observed at the $3\sigma$ level.  LBNE10 will have $3\sigma$ sensitivity for a significant fraction of $\delta_{CP}$ values if $\sin^2\theta_{23}$ is in the lower octant. LBNE will have significant sensitivity for all $\sin^2\theta_{23}$ values.
The right panel of Fig. \ref{fig:cpv} shows the sensitivity, in $\sigma$, to which LBNE10 and LBNE can detect CP violation as function of the true value of $\delta_{CP}$, for the case in which $\sin^2\theta_{23}$ is equal to the global best-fit value\cite{Fogli:2012ua}. CP violation can be detected at the $3\sigma$ level for 40\% of $\delta_{CP}$ values with LBNE10 and 67\% of $\delta_{CP}$ values with LBNE.

\begin{figure}[!h]
  \centering
  \includegraphics[width=0.4\textwidth,trim=0.14in 0in 0.4in 0.4in,clip]{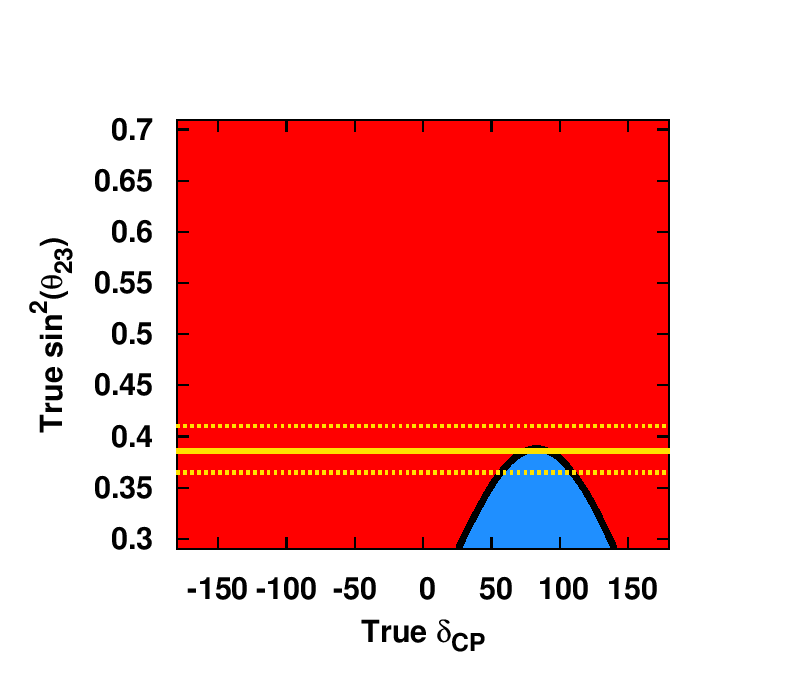}
  \includegraphics[width=0.4\textwidth,trim=0in 0in 0in 0.4in,clip]{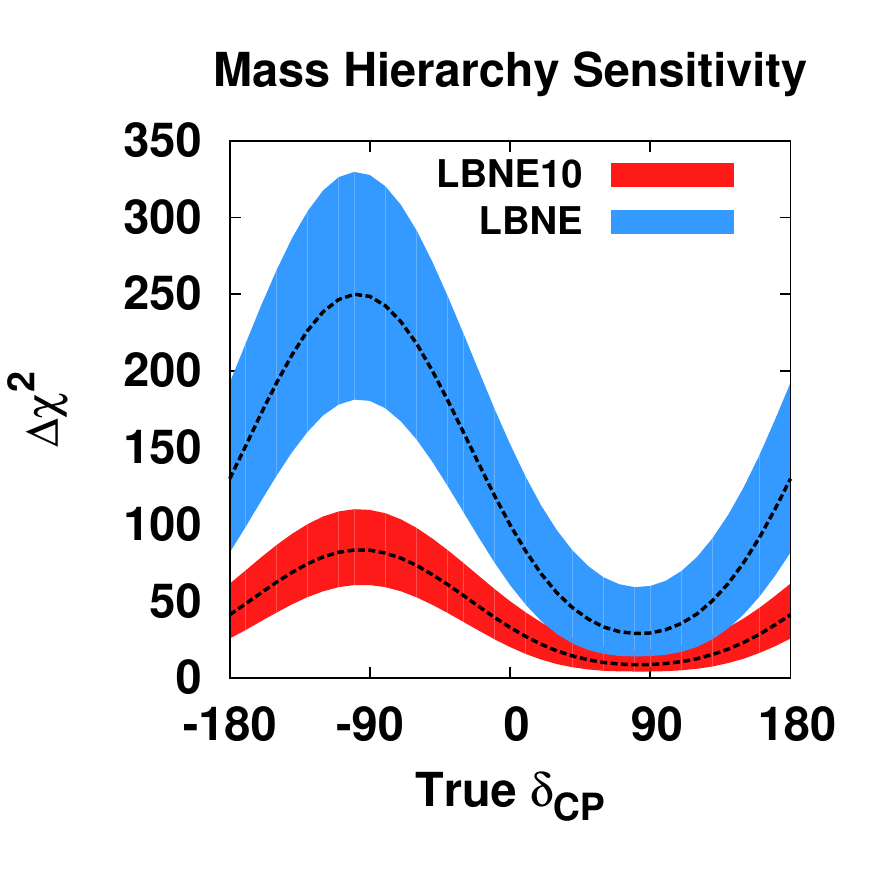}
  \caption{Mass hierarchy sensitivity assuming true normal hierarchy. (left) Regions in $\sin^2\theta_{23}$ vs. $\delta_{CP}$ where LBNE (red and cyan) and LBNE10 (red) have $\Delta \chi^2 \geq9$ for the normal mass hierarchy. The central value($1\sigma$ range) on $\sin^2\theta_{23}$ from the Fogli et al. 2012 global fit are indicated with the solid(dashed) yellow lines. (right) Expected sensitivity, in $\Delta \chi^2$, with which LBNE and LBNE10 will be able to determine the mass hierarchy at $\sin^2\theta_{23}=0.39$. The bands represent current uncertainties on $\theta_{23}$ and $\Delta m^2_{31}$ and projected uncertainties on $\theta_{13}$.}
  \label{fig:mh}
\end{figure}

Figure \ref{fig:mh} shows the performance, in terms of $\Delta \chi^2$\footnote{Qian et al. \cite{Qian:2012zn} have shown that the probability of correct mass hierarchy determination is not correctly described by frequentist statistical methods, and provide a method for properly calculating them based on these results.}, for LBNE10 and LBNE in determining the neutrino mass hierarchy. The $\Delta \chi^2\geq9$ regions where the mass hierarchy can be determined show nearly complete coverage for LBNE10 and complete coverage for LBNE.
The sensitivity as a function of $\delta_{CP}$ shows that LBNE10 can determine the mass hierarchy at $\Delta \chi^2\geq9$ for most $\delta_{CP}$ values while LBNE has nearly complete $\delta_{CP}$ coverage for mass hierarchy determination with $\Delta \chi^2\geq25$.

Studies, not shown here, were also done to assess the sensitivity added by including information from T2K and NO$\nu$A. In LBNE10, a combined fit to LBNE10, T2K, and NO$\nu$A provides a significant boost to the sensitivity. For example, for the worst case mass hierarchy sensitivity (around $\delta_{CP}=90^\circ$ in the normal hierarchy) the $\Delta \chi^2$ improves from $\sim9$ to $\sim16$. However, for high exposures in LBNE, the combined, T2K+NO$\nu$A+LBNE fit contributes little over LBNE alone to the sensitivity. 

\section{Conclusion}
LBNE will be a major step forward in ability to constrain the neutrino oscillation parameters of the PMNS matrix. Even early stages of the experiment, characterized here by LBNE10, represent a large improvement on current and near-future neutrino oscillation measurements. With the assumptions outlined here, LBNE will be able to measure $\delta_{CP}$ to between $10^\circ$ and $20^\circ$, detect CP violation at $3\sigma$ for 67\% of $\delta_{CP}$ values, and resolve the neutrino mass hierarchy with $\Delta \chi^2\geq25$.

\end{document}